\documentclass[aps,groupedaddress,showpacs,showkeys]{revtex4-1}
\usepackage[utf8]{inputenc}
\usepackage{verbatim}
\usepackage{natbib}
\usepackage{lipsum}
\usepackage{mathtools}
\usepackage{amsmath}
\usepackage{amsfonts}
\usepackage{amssymb}
\usepackage{epsfig}
\usepackage{wrapfig}
\usepackage{changepage}
\usepackage{float}
\usepackage{wrapfig}
\usepackage{breqn}
\usepackage{needspace}

\newcommand{\beq}{\begin{equation}}
\newcommand{\eeq}{\end{equation}}
\begin{document}

\title{Thermodynamic force thresholds biomolecular behavior}

\makeatletter
\let\cat@comma@active\@empty
\makeatother
\author{Milo M. Lin}
\thanks{Milo.Lin@UTSouthwestern.edu}
\affiliation{Green Center for Systems Biology \\ Department of Biophysics \\ Center for Alzheimer's and Neurodegenerative Diseases \\ University of Texas Southwestern Medical Center, Dallas, Texas, 75235.}


\begin{abstract}

 In living systems, collective molecular behavior is driven by thermodynamic forces in the form of chemical gradients. Leveraging recent advances in the field of nonequilibrium physics, I show that increasing the thermodynamic force alone can induce qualitatively new behavior. To demonstrate this principle, general equations governing kinetic proofreading and microtubule assembly are derived. These equations show that new capabilities, including catalytic regulation of steady-state behavior and exponential enhancement of molecular discrimination, are only possible if the system is driven sufficiently far from equilibrium, and can emerge sharply at a threshold force. Regardless of design parameters, these results reveal that the thermodynamic force sets fundamental performance limits on tuning sensitivity, error, and waste. Experimental data show that these biomolecular processes operate at the limits allowed by theory.
\end{abstract}

\maketitle
\noindent\textbf{INTRODUCTION}
\newline
Thermodynamic forces, in the form of chemical potential differences, drive biomolecular self-assembly, directed motion, and signaling. It is generally understood that these forces, which expend energy, are needed to achieve capabilities that would be forbidden at equilibrium. However, the quantitative relationship between thermodynamic force and non-equilibrium capabilities is not well understood. For example, seminal work established upper limits achievable by equilibrium systems in terms of sensitivity \cite{hopfield} and sharpness \cite{gunawardena_sharpness,depace_sharpness}, and demonstrated theoretically and experimentally that introducing thermodynamic forces can enhance these capabilities beyond equilibrium limits. However, much less is known about how much thermodynamic force is required to achieve these capabilities. Some intriguing relations between kinetic proofreading error and entropy production rate, which is an emergent property that relates to thermodynamic force, have been found numerically and in the asymptotic limit \cite{gunawardena_proofreading_final}. Yet, overall progress is limited because obtaining interpretable general insights into complex non-equilibrium systems is mathematically challenging \cite{hill,schnakenberg,zia}. 
\par
The main objective of complexity reduction is to redefine systems in terms of new collective variables that are conducive to mathematical simplification and abstraction \cite{Anderson}. The collective variables would ideally encode the properties that emerge from the interactions, thereby explaining how the system behaves differently than the sum of its parts. Recent advances from the field of non-equilibrium statistical physics show that probability flux between the states of a dynamical system is mathematically equivalent to Ohm's law if mapped to the appropriate circuit representation \cite{circuit}. This mapping provides a quantitative framework for systematically modularizing complex systems to reveal biomolecular design principles. Of particular interest for biomolecular thermodynamics, this framework introduces collective variables that isolate the effect of the thermodynamic force on system behavior \cite{circuit}. Leveraging this framework, I show that emergent properties can appear if the thermodynamic force (i.e. concentration of chemical fuel) exceeds a threshold value. For experimentally characterized processes of microtubule assembly and kinetic proofreading, I show how two foundational capabilities, which are impossible at equilibrium, emerge only at high force: (i) catalytic control of steady-state behavior and (ii) exponential enhancement of binding sensitivity. These analytic results are not dependent on the detailed system parameters, but can emerge sharply as a function of thermodynamic force. Experimentally-observed relationships which were not intuitively interpretable are revealed to be the result of biological systems operating at upper limits allowed by the theory. These design principles shed light on how the toolkit of biomolecular operations is appreciably expanded in the presence of large chemical gradients. 
\newline

\noindent\textbf{RESULTS}
\newline
\textbf{Circuit framework for simplifying biomolecular processes.} 
Recently, we identified a mathematical equivalence between Markovian dynamical systems and electronic circuits. The elementary "direct" variables of free energies and rate constants are transformed into new variables corresponding to circuit elements: resistors and batteries \cite{circuit}. The system is driven out of equilibrium by thermodynamic forces (mapped to "batteries") that maintain high concentrations of "energy currency" molecules such as ATP and GTP, typically many orders of magnitude above their equilibrium concentrations. Thermodynamic force is the chemical potential difference, which is in units of energy per particle rather than energy per distance, and is given in the natural units of thermal energy (kT) per particle. The magnitude of the thermodynamic force is therefore a measure of how far the system is from equilibrium. To solve the steady-state probability distribution of a biomolecular system involves calculating the probability fluxes that satisfy an Ohm's law relation called the probability flow equation (PFE). As in electronic circuits, the resistors in a probability circuit can be systematically combined into collective variables by merging them in parallel or in series \cite{EEbook}. Using these rules, a complex system can be simplified to a minimal set of irreducible resistors that are explicit functions of the microscopic parameters, and the thermodynamic force \cite{circuit}. In the following two sections, I apply this circuit framework to solve the PFE of two fundamental biomolecular processes: microtubule self-assembly, and error-correction in translation. Doing so yields simple closed-form steady-state solutions of system output that explicitly separates the role of the thermodynamic force from the system design parameters. 

\textbf{Catalytic control of microtubule length.}
Biomolecular systems are capable of catalytic control, defined as the ability of a catalyst to change the steady-state properties of a system. This property, which is not possible at equilibrium, allows a catalyst molecule to exert an influence at sub-stoichiometric concentrations relative to its target substrate, significantly relieving spatial constraints within the crowded cell environment. 
\Needspace{50pt}
\begin{wrapfigure}{l}{0.44\textwidth}
    \begin{center}
     \includegraphics[width= 0.44 \textwidth]{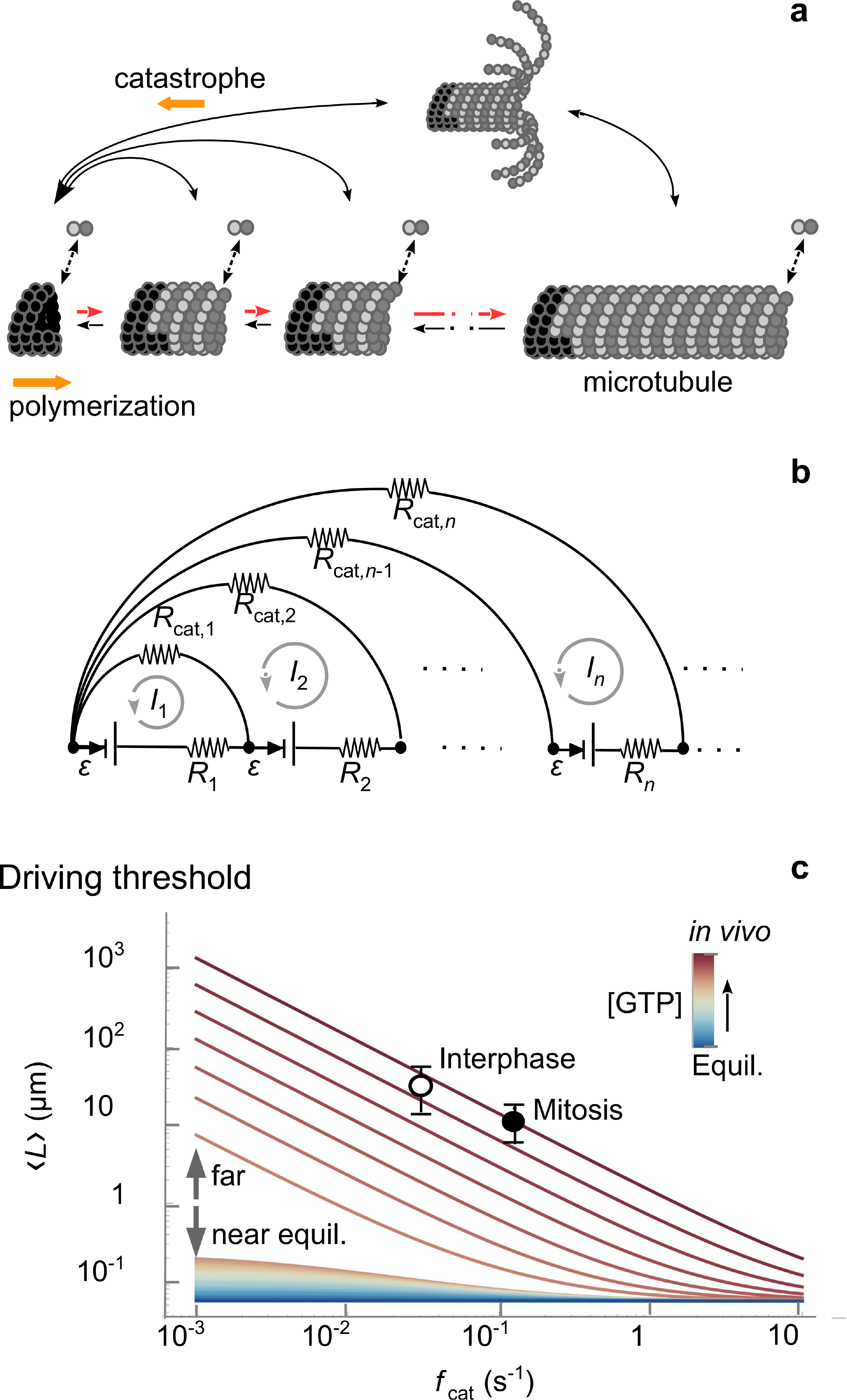}
    \end{center}
    \caption{\textsf{\textbf{Catalytic regulation of microtubule length} Self-assembly of tubulin subunits into filaments, with complete dis-assembly events (catastrophes) (a). This process can be mapped to a circuit diagram (b). Eq. 1 gives the mean length as a function of catastrophe frequency and thermodynamic force, which can be interpreted as the GTP concentration (color bar; c). At equilibrium (blue line), catastrophe frequency has no influence on mean length. At physiological GTP concentration (red), the predicted frequency dependence is in excellent agreement with the measured mean microtubule length at interphase and mitosis, which differ only in the catastrophe frequency (c). The catastrophe frequency linearly tunes mean microtubule length above a critical thermodynamic force (GTP concentration). Parameters and measured lengths are taken from Ref. 21. }}
    \label{fig1}
\end{wrapfigure}
Consequently, it is utilized in a broad range of contexts in which protein activity or assembly are tuned by multiple regulatory proteins. For example, some kinases are more than three orders of magnitude lower in concentration than their target substrates \cite{kinase_concentration}. 
Here, I consider the regulation of microtubule length by catalytic regulators that (de)stabilize the growing microtubule end cap. The elementary steps constituting microtubule self-assembly is shown in Fig.1A in the absence of rescue from catastrophe \cite{dynamic_instability}; the state space contains an infinite number of possible states in the thermodynamic limit. The reversible assembly of GTP-bound tubulin dimers occurs with forward and reverse rate constants $k_f$ and $k_r$ if GTP were allowed to equilibrate with GDP. Thus, $k_f/k_r=e^{-\beta G}$, where $G$ is the equilibrium dimer binding free energy. 
In cells, GTP is kept at high concentration in excess of its equilibrium level, effectively giving rise to an additional forward rate constant $\alpha$, which is proportional to excess [GTP] up to a saturation concentration. The thermodynamic force $\Delta \mu = kT \ln{[1+\alpha/k_f]}$; for mathematical clarity, in the following $\alpha$ (and [GTP]) will serve as a measure of thermodynamic force. Assembly is counteracted by the catastrophe rate constant $f_{\mathrm{cat}}$, which leads to complete dis-assembly of the microtubule in a regulatable manner \cite{luke2}. Catastrophe is triggered by the stochastic disruption of the growing microtubule cap \cite{microtubule_capping}, which allows cap-modifying substrates to act as sub-stoichimetric catalysts of microtubule shrinkage. The dynamic instability steady state is reached when the catastrophe balances net dimer addition, in contrast to the detailed balance steady state of an equilibrium system \cite{dynamic_instability}. Although this process has been modeled mathematically \cite{microtubule_langevin} and via computational simulations\cite{microtubule_simulations}, the complexity of the dynamical system consisting of numerous reversible reactions, has limited our quantitative understanding of how system parameters control microtubule length distributions. Previous work has established the intrinsic speed-up of non-equilibrium polymer reorganization kinetics compared to equilibrium reorganization \cite{nonequilibrium_polymerspeed}, yet the advantages of nonequilibrium catastrophe-based regulation on steady-state observables (e.g. microtubule length) is poorly understood.
\par
Equilibrium theory teaches that catalytic rate constants cannot affect the mean value of any observable. In contrast, the microtubule length probability distribution $P(L)$ reaches a bounded steady state with well-known mean length explicitly dependent on $f_{\mathrm{cat}}$: $<L> = (\alpha+k_{\mathrm{f}}-k_{\mathrm{r}})/f_{\mathrm{cat}}$ \cite{microtubule_leibler}. 
In this regime, a catalyst which only decreases the energy barrier to catastrophe leads to a proportional change in the mean length, in violation of the equilibrium rule. This catalytic regulation in fact occurs during the eukaryotic cell cycle, where increased $f_{\mathrm{cat}}$ causes the decrease in microtubule length necessary for cell division \cite{mitchison}. Yet, the thermodynamic force required to enable catalytic regulation has remained unclear.
\par
I mapped this process to the circuit framework (Fig. 1B; $\mathcal{E} \propto \alpha/k_f$) to obtain the closed-form expression for the steady state length distribution of microtubules (See Appendix):

\begin{equation}
P(L)= {{f_{\mathrm{cat}}}\over{f_{\mathrm{cat}}-\alpha (e^{\beta G}-1)}}P_1e^{-\beta G(L-1)}+{{\alpha(e^{\beta G}-1)}\over{\alpha (e^{\beta G}-1)-f_{\mathrm{cat}}}}P_1 e^{-D(L-1)},
\end{equation}
\noindent
\Needspace{40pt}
\begin{wrapfigure}{l}{0.44\textwidth}
    \begin{center}
     \includegraphics[width= 0.4 \textwidth]{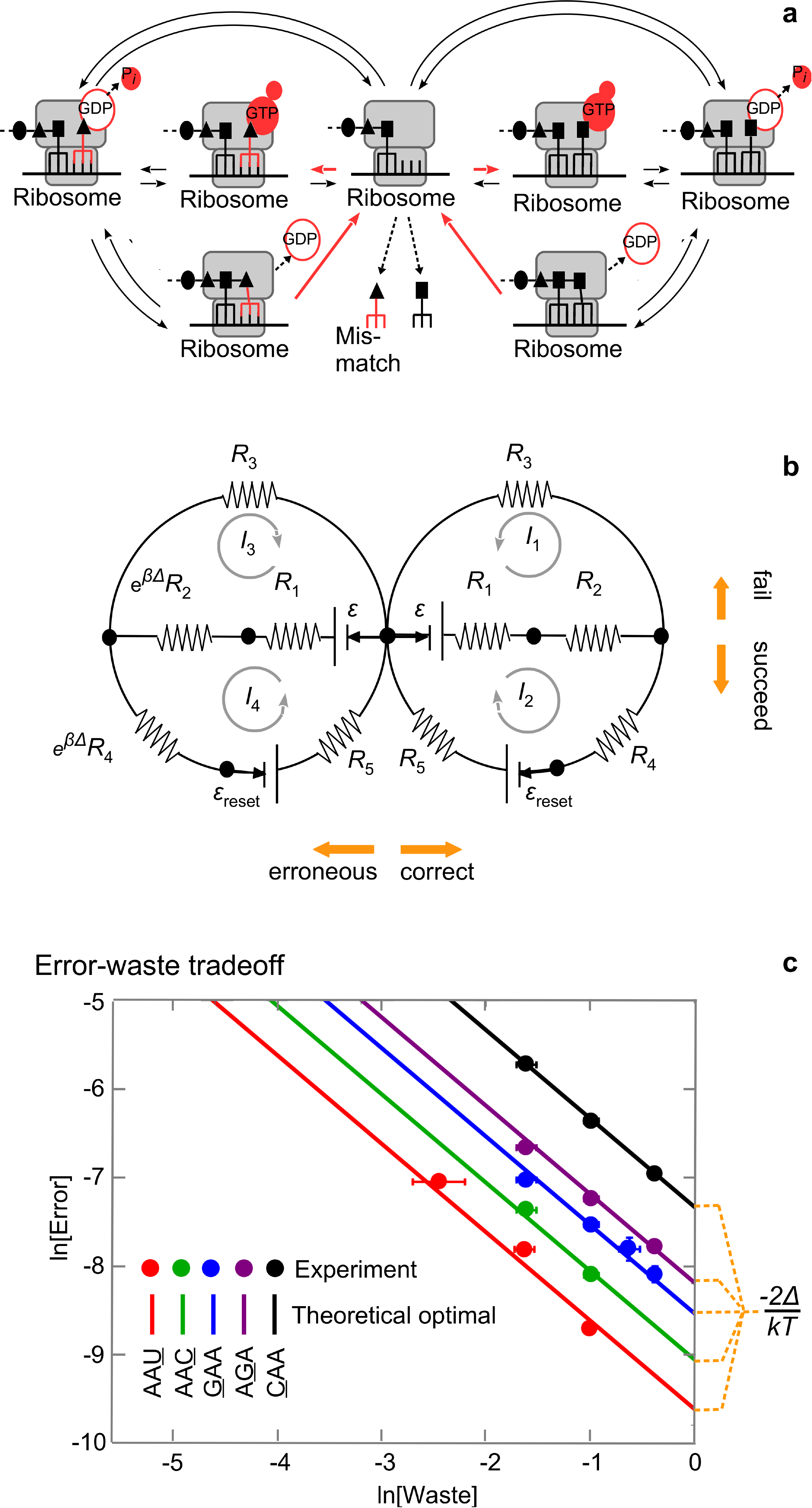}
    \end{center}
    \caption{\textsf{\textbf {Error and waste tradeoff in kinetic proofreading.}  Formation of the activated complex follows the correct or erroneous codon binding (less energetically favorable by $\Delta$), which can succeed or fail (a). The currents in the circuit mapping (b) can be solved analytically at steady state to give Eq. 3, which predicts optimal tradeoff between error and waste (c). Experimental data (colored circles, taken from Ref. 25) for the error correction of five point mutations of the codon AAA shows that this system operates at the optimal tradeoff predicted by the theory, with each mutation corresponding to a different $\Delta$ (colored lines). For each point mutation, the error-waste tradeoff is tuned by varying magnesium concentration.}}
    \label{fig2}
\end{wrapfigure}
\noindent where $P_1 \equiv P(1)$ is the monomer fraction, and $D \equiv - \ln{[1-{{\sqrt{(\alpha+f_{\mathrm{cat}}+k_f-kr)^2+4f_{\mathrm{cat}}k_r}-(\alpha+f_{\mathrm{cat}}+k_f-kr)}\over{2k_r}}]}$. Although mean filament length has been calculated using generating functions \cite{microtubule_length}, this is the first time that $P(L)$ has been solved and the role of the thermodynamic force isolated. Interestingly, $P(L)$ is  a superposition of two exponential functions, corresponding to the equilibrium and nonequilibrium contributions, respectively. The double exponential explains why previous attempts to fit $P(L)$ generated from numerical simulations to a single exponential distribution led to poor fits\cite{microtubule_length}.  
Fig.1c shows the mean microtubule length as a function of catastrophe rate as predicted by Eq. 1 using measured rate constants \cite{microtubule_howard}, for varying $\alpha$ corresponding to different GTP concentrations. As expected, Eq. 1 reduces to the equilibrium single-exponential distribution, which is independent of $f_{\mathrm{cat}}$, if $\alpha = 0$ (blue line). Eq. 1 predicts that, as the system is driven from equilibrium, the length distribution jumps between two distinct regimes with qualitatively different dependence on $f_{\mathrm{cat}}$. The jump occurs when $\alpha$ exceeds $k_r - k_f$. In the strongly-driven regime, for which $(\alpha+k_{\mathrm{f}}-k_r)/f_{\mathrm{cat}} >> 1$, Eq.1 simplifies to $<L>_{\mathrm{strong}} = (\alpha+k_{\mathrm{f}}-k_{\mathrm{r}})/f_{\mathrm{cat}}$, which is the well-known formula cited above; at physiological GTP concentrations, the predicted mean length is in excellent agreement with measured lengths \cite{mitchison} in both Mitosis and Interphase (circles in Fig.1c). In the weakly-driven regime ($-(\alpha+k_{\mathrm{f}}-k_r)/f_{\mathrm{cat}} >> 1$), Eq.1 simplifies to $<L>_{\mathrm{weak}}=  -\ln{[{{\alpha+k_f}\over{kr}}+{{f_{\mathrm{cat}}(\alpha+k_f)}\over{kr(\alpha+k_f-k_r)}}]}^{-1}$; the mean length is only marginally sensitive to $f_{\mathrm{cat}}$ in this regime. The thermodynamic force, as parameteried by $\alpha$ or [GTP], controls the transition between the near and far-from-equilibrium regimes, whose sharpness is inversely proportional to $f_{\mathrm{cat}}$(Fig.1C and Appendix). Therefore, a uniquely non-equilibrium feature (catalytic regulation of an ensemble-averaged observable) is switched on in an all-or-none fashion when the system is driven beyond the threshold level.  
\par

\textbf{Error-waste tradeoff in kinetic proofreading.}
It has long been appreciated that by expending energy, molecular discrimination can be enhanced beyond the constraints of chemistry \cite{hopfield}. For example, cells perform energy-intensive error-correction in order to tolerate errors made in translating the genetic code into proteins. The correct and erroneous tRNA match can both bind to messenger RNA, with erroneous binding being $\Delta$ less energetically stable than the correct match (Fig. 2A). Following binding, the complex is driven to the activated state via binding of GTP, corresponding to the battery potential $\mathcal{E}$, after which protein elongation can proceed to completion or fail. $\mathcal{E} \propto \alpha/k_1$, which is the enhancement of the rate constant of GTP binding divided by the equilibrium rate constant. Because the chemical potential difference $\Delta \mu = \ln{[1+\alpha/k_1]}$ \cite{circuit}, I will use $\alpha/k_1$ as a measure of thermodynamic force. 
\par
Upon successful completion (whether correct or erroneous), the process is then reset back to the original state via $\mathcal{E}_{\mathrm{reset}}$, corresponding to the maintenance of fixed product and reactant concentrations. In his classic paper, Hopfield showed that the minimum error is $e^{-2\beta \Delta}$ \cite{hopfield}. However, it has long been appreciated that this limit is incomplete because of other performance characteristics that must constrain the system. For example, limits on the speed as well as the error have been derived for proofreading regimes of varying complexity \cite{error_speed, murugan}. More recently, the importance of energy efficiency, and therefore the reduction of wasteful cycles, has been quantitatively demonstrated to be a limiting constraint for bacterial fitness \cite{error_fitness}, and an intriguing linear trade-off between accuracy and efficiency has been empirically observed \cite{linear_tradeoff}. In this work, the efficiency is quantitatively defined by dividing the speed of correct product formation by the weighted sum of the speeds of all processes, including erroneous or failed events. The waste is one minus the efficiency. A constraint on the error, speed, and entropy production rate was derived in the asymptotic limit, with an intriguing tighter bound found by numerical simulation \cite{gunawardena_proofreading_final}. However, the direct tradeoff between error and waste remains an open question. Furthermore, how this tradeoff is conditional on the thermodynamic force is poorly understood.
\par
I use the circuit framework to establish the direct tradeoff between error and waste, and show that there is a qualitatively difference between the low force and high force regimes. The circuit diagram for this process is shown in Fig. 2b. The speed of correct product formation is $I_2$, error $\epsilon=I_4/I_2$, efficiency $\eta = I2/(I1+I2+I3+I4)$, and waste $\omega = 1- \eta$. The steady-state PFE over the four simple loops of the circuit can be analytically solved after summing $R_1$ and $R_2$ in series, to express the error as a function of waste and resistors (See Fig. 2b and Appendix).
\par
In the low-force regime ($\alpha << k_1$), the system is near equilibrium and the expression simplifies to:
 \begin{equation}
\epsilon_{\mathrm{weak}} >\bigg(1-{{\alpha/k_1}\over{1-\omega}}\bigg) e^{{-\Delta}\over{k_{\mathrm{B}}T}},
\end{equation}
which gives the equilibrium bound of $e^{{-\Delta}\over{k_{\mathrm{B}}T}}$ when $\alpha = 0$ \cite{hopfield}. Note that Eq. 2 applies under the condition that $\alpha/k_1 < 1-\omega$; therefore, in the low force regime reducing waste is not a significant constraint to achieving the optimal error.
\par
In the biologically-relevant high force (i.e. far-from-equilibrium) regime in which $\alpha >> k_1$, the expression simplifies to a tight lower bound on the error $\epsilon$ as a function of the waste $w$ that is qualitatively different from the low force relation: 
 \begin{equation}
\epsilon_{\mathrm{strong}} >{{1}\over{\omega}} e^{{-2\Delta}\over{k_{\mathrm{B}}T}}.
\end{equation}
  This result generalizes Hopfield's relation by integrating the additional constraint of waste. It immediately indicates that Hopfield's minimum error bound can only be achieved in the limit of maximum waste (i.e. zero efficiency). In fact, Eq. 3 can be reformulated to emphasize, mathematically, that error and waste have an equal trade-off that is capped by the binding discrimination between the correct and incorrect codons: $\ln\epsilon + \ln\omega  > {{-2\Delta}\over{k_{\mathrm{B}}T}}$. Note that this simple tight inequality is obtained by considering all possible values of the resistors in the equality obtained from solving the PFE (See Appendix). Therefore, the only parameter that influences the optimal trade-off is $\Delta/k_{\mathrm{B}}T$. Eq. 3 quantitatively explains the empirically-observed linear tradeoff between efficiency and accuracy \cite{linear_tradeoff}. Fig. 2c demonstrates this, showing experimental data for the translational error and waste of the codon AAA for five different single-nucleotide mutations \cite{linear_tradeoff}; the protein translation machinery achieves the optimal bound allowed by Eq. 3 as it trades error for waste under different magnesium ion concentrations. Therefore, even if the system can optimize over all possible parameters, waste reduction switches from a "soft" sub-leading constraint to a "hard" leading-order constraint with increasing thermodynamic force.
\newline

\noindent\textbf{DISCUSSION}
\newline
The main result of this work is to mathematically demonstrate that qualitatively new properties can emerge, sometimes sharply, for interacting systems if they are coupled to a sufficiently strong energy gradient. New capabilities, and their limits, were shown to arise for systems driven by a sufficiently large thermodynamic force, and experimental data show that biological processes perform very close to these limits.  These findings may have implications for the necessity of establishing a large driving gradient (such as ATP and GTP concentrations) as a hard prerequisite to subsequent optimization or evolution. As examples, this work mostly focuses on molecular biology systems with high-resolution experimental measurements. It is reasonable to expect that a nonequilibrium switch applies much more generally, and that perhaps this switch could help delineate the categorical difference between living and nonliving systems at the molecular scale.
\newline

\noindent\textbf{ACKNOWLEDGEMENTS}
\newline
\noindent The author would like to thank Elliot Ross, Luke Rice, and Alan Katz for valuable feedback on the manuscript. This work was supported by the Cecil and Ida Green Foundation.

\bibliographystyle{pnas}
\bibliography{Nonequil}

\newpage
\begin{widetext}
\section{Appendices}
\subsection{The circuit framework and the probability flow equation (PFE)}

The mapping from the master equation of a Markovian dynamical system to a circuit that obeys the probability flow equation (PFE), which is in the form of Ohm's law, is derived in detail in Ref. 9. Here, the PFE and the mapping from thermodynamic and kinetic parameters to circuit variables is briefly summarized (See Fig. S1). 
\setcounter{figure}{0}
\renewcommand{\thefigure}{S\arabic{figure}}
\begin{figure} [H]
\centering
\includegraphics[width= 0.85 \linewidth]{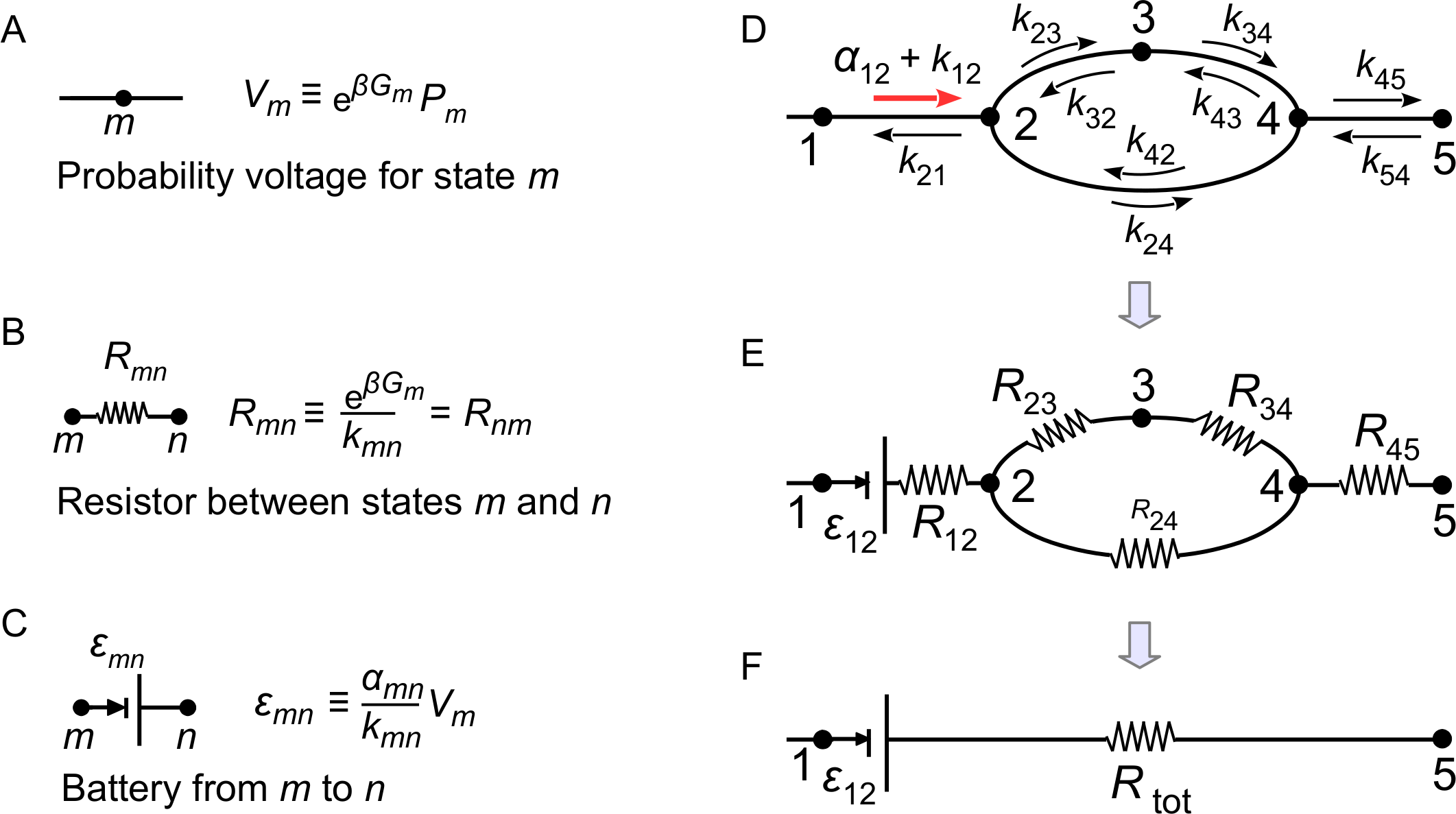}
\caption{\textsf{Circuit framework. The probability potential $V$ is a state variable defined for each state of the system (A). Resistors (B) are passive circuit elements that only depend on the equilibrium parameters of the system decoupled from any thermodynamic forces such as chemical gradients. Batteries (C) correspond to active circuit elements that directly couple thermodynamic forces to the system, which is reflected in an excess non-equilibrium rate constant (red arrow in D), which can be directly converted to the chemical potential difference driving the system. Using these circuit variables, any dynamical system master equation (D) can be mapped to a circuit obeying Ohm's law (E), and simplified using the tools from the field of electronic circuit theory (F, see Ref. 9). }}
\label{figs1}
\end{figure}

A dynamical system consists of states and rates of transitions between the states. In treating biologically relevant systems, we decompose the system into an underlying "reference" equilibrium system that is coupled to energy gradients (thermodynamic forces) that drive the system away from equilibrium. If the system is ergodic, we can consider the steady-state probability of the system occupying each of its states. For a particular state $i$, denote the steady-state probability $P_i$. The state $m$ is also associated with a reference free energy $G_i$, which is the free energy of state $i$ when the energy gradient is zero. In this case, $P_i$ would be the probability of finding the equilibrium system in state $i$, and is given by the well-known Boltzmann distribution:
\begin{equation}
P_i ^* = {{\mathrm{e}^{-\beta G_i}} \over {Z}},
\end{equation} 
where the asterisk denotes the special case of an equilibrium system, $G_i$ is the free energy of state $i$, $\beta = 1/kT$ is one over temperature times Boltzmann's constant, and the partition function $Z$ ensures probability normalization: $Z = \sum_k {\mathrm{e}^{-\beta G_k}}$. We can write the same relation for any other state $j$, and combining with the above to recover the oftentimes more useful version of the Boltzmann distribution relating the relative probabilities of any two states of an equilibrium system to the difference in their free energies:
\begin{equation}
P_i^* = P_j^* {\mathrm{e}^{\beta (G_j-G_i)}}.
\end{equation} 

The system will be able to transition directly from a given state $m$ to an adjacent state $n$, with the forward rate given by mass action: $P_m k_{mn}$, where $k_{mn}$ is the rate constant of transitioning from state $m$ to state $n$. In the circuit mapping, the "resistance," $R_{mn}$, between states $m$ and $n$ is defined:
 \begin{equation}
R_{mn} \equiv {{\mathrm{e}^{\beta G_m}} \over {k_{mn}}}=R_{nm}.
\end{equation} 

The resistance is a function only of the parameters of the reference equilibrium system. Note that the second equality above follows because the equilibrium forward transmittance is equal to the equilibrium backward transmittance; hence the resistance is directionally symmetric, just like the behavior of standard resistors in electronic circuits. This property motivates the mapping of the other terms to elements of a circuit. Define the probability "potential" of a state to be:
 \begin{equation}
V_{m} \equiv P_{m} \mathrm{e}^{\beta G_m}.
\end{equation} 
Intuitively, the potential of a state is (up to normalization by the partition function) its probability divided by its equilibrium probability - a driving force for probability flows. If there is a mass source or sink in the system, which is a state whose probability (i.e. potential) remains unchanged by probability flow into or out of the state, such states correspond to "grounds." Finally, define the "battery" driving transitions from $m$ to $n$ as:
 \begin{equation}
\mathcal{E}_{mn} \equiv {\alpha_{mn} \over {k_{mn}}} \mathrm{e}^{\beta G_m}P_{m},
\end{equation} 
which is proportional to the driven rate constant divided by the equilibrium rate constant and is zero when the transition between $m$ and $n$ is not driven. $\mathcal{E}_{mn}$ is also proportional to the potential at $m$; the battery is therefore a potential-feedback battery. 
\par
Using these definitions, the probability flow equation along any path between two states $i$ and $j$ is determined by the potential difference between the states: 
\begin{equation}
V_{j} -V_{i} = {\sum\limits_{m=i}^{n=j}(\mathcal{E}_{mn}-R_{mn}I_{mn})},
\end{equation}
where the sum is over \textit{any} path from state $i$ to state $j$ parameterized by neighboring states $m$ and $n$ along such a path. In terms of the dynamical system parameters, the PFE is:

\begin{equation}
P_{j} \mathrm{e}^{\beta G_j}  - P_{i} \mathrm{e}^{\beta G_i} = \sum\limits_{m=i}^{n=j}\Big({\alpha_{mn} \over {k_{mn}}} P_{m} \mathrm{e}^{\beta G_m} -{{{\mathrm{e}^{\beta G_m}} \over {k_{mn}}}}I_{mn}\Big).
\end{equation} 
As expected, if the sytem is not driven, then all $\alpha$ are zero and all $I$ are zero (no net currents), and the PFE reduces to the Boltzmann distribution (Eq. 5).
\subsection{Microtubule self-assembly}
In terms of the $n$th mesh current shown in Fig. 1b, the voltage equation taken along the path of the $n$th battery is:
\begin{equation}
P_{n+1}e^{\beta G_{n+1}}-P_{n}e^{\beta G_n}= {{\alpha}\over{k_f}}P_n e^{\beta G_n} - R_n I_n.
\end{equation}
where $R_n = e^{\beta G_n}/k_f$ and $G_n=nG$. Note that $e^{\beta G} = k_b/k_f$, where $k_f$ and $k_b$ are the equilibrium forward and backward rates, respectively. Using these definitions, we can solve for the probability of the $(n+1)$th state in terms of the previous state probability and current:
\begin{equation}
P_{n+1}=\Big(1+{{\alpha}\over{k_f}} \Big) e^{-\beta G}P_n-I_n {{e^{-\beta G}}\over{k_f}}
\end{equation}

Taking the potential difference from state $n+1$ and state 1 along the catastrophe path:
\begin{equation}
P_{1}e^{\beta G_{1}}-P_{n+1}e^{\beta G_n+1}= -R_{\mathrm{cat}, n} (I_{n}-I_{n+1}),
\end{equation}
Where $R_{\mathrm{cat}, n} = e^{\beta G_{n+1}}/f_{\mathrm{cat}} = R_{n+1} ({{k_f}\over{f_{\mathrm{cat}}}})$.
Therefore, the $(n+1)$th current is:
\begin{equation}
I_{n+1}= I_n - f_{\mathrm{cat}}P_{n+1}+ f_{\mathrm{cat}}P_{1}e^{-\beta G_{n}}
\end{equation}

In vector notation, the recursive probability and current equations become:
\begin{equation}
\begin{bmatrix}
    1       & 0 \\
    f_{\mathrm{cat}}       & 1
\end{bmatrix}
\begin{bmatrix}
    P_{n+1} \\
    I_{n+1} 
\end{bmatrix}
=
\begin{bmatrix}
    \Big(1+{{\alpha}\over{k_f}} \Big) e^{-\beta G}       & -{{e^{-\beta G}}\over{k_f}} \\
    0      & 1
\end{bmatrix}
\begin{bmatrix}
    P_{n} \\
    I_{n} 
\end{bmatrix}
+
\begin{bmatrix}
    0 \\
    {{f_{\mathrm{cat}}P_{1}}\over{e^{\beta G_{n}}}}
\end{bmatrix}
\end{equation}
Multiplying both sides by the inverse of the right-hand-side matrix, the recursion relation is:
\begin{equation}
\begin{bmatrix}
    P_{n+1} \\
    I_{n+1} 
\end{bmatrix}
=
M
\begin{bmatrix}
    P_{n} \\
    I_{n} 
\end{bmatrix}
+
\begin{bmatrix}
    0 \\
    {{f_{\mathrm{cat}}P_{1}}\over{e^{\beta G_{n}}}}
\end{bmatrix}
\end{equation}
where the transition matrix $M$ is given by:
\begin{equation}
M=
\begin{bmatrix}
    \Big(1+{{\alpha}\over{k_f}} \Big) e^{-\beta G}       & -{{e^{-\beta G}}\over{k_f}} \\
    -f_{\mathrm{cat}}\Big(1+{{\alpha}\over{k_f}} \Big) e^{-\beta G}      & f_{\mathrm{cat}}{{e^{-\beta G}}\over{k_f}}+1
\end{bmatrix}
\end{equation}

The probability and current of state $n$ in terms of those of state 1 is thus:
\begin{equation}
\begin{bmatrix}
    P_{n+1} \\
    I_{n+1} 
\end{bmatrix}
=
M^n
\begin{bmatrix}
    P_{1} \\
    I_{1} 
\end{bmatrix}
+\sum_{k=0}^{n-1}(e^{\beta G}M)^k
\begin{bmatrix}
    0 \\
    {{f_{\mathrm{cat}}P_{1}}{e^{-\beta G n}}}
\end{bmatrix}
\end{equation}
Diagonalizing $M$:
\begin{equation}
M=V
\begin{bmatrix}
    \lambda_-       & 0 \\
    0      & \lambda_+
\end{bmatrix}
V^{-1}
\end{equation}
Where the columns of $V$ are the eigenvectors of $M$ and $\lambda_-$ and $\lambda_+$ are the eigenvalues of $M$:
\begin{equation}
\lambda_{\pm} = {{e^{-\beta G}}\over{2k_f}} \Bigg( \alpha+ k_f(1+e^{\beta G})+f_{\mathrm{cat}} \pm \sqrt{(\alpha+k_f-k_fe^{\beta G})^2+f_{\mathrm{cat}}(2\alpha+2(1+e^{\beta G})k_f+f_{\mathrm{cat}})}\Bigg)
\end{equation}
Which simplifies to the value given in the text:
\begin{equation}
D=-\ln{\lambda_{-}} = -\ln{[1- {{\sqrt{(\alpha+f_{\mathrm{cat}}+k_f-kr)^2+4f_{\mathrm{cat}}k_r}-(\alpha+f_{\mathrm{cat}}+k_f-kr)}\over{2k_r}}]} 
\end{equation}

Note that $\lambda_- \leq 1$ whereas $\lambda_+ \geq 1$.

The transfer matrix equation is then
\begin{equation}
\begin{bmatrix}
    P_{n+1} \\
    I_{n+1} 
\end{bmatrix}
=
V
\begin{bmatrix}
    \lambda_-^n       & 0 \\
    0      & \lambda_+^n
\end{bmatrix}
V^{-1}
\begin{bmatrix}
    P_{1} \\
    I_{1} 
\end{bmatrix}
+\sum_{k=0}^{n-1}e^{\beta k G}
V
\begin{bmatrix}
    \lambda_-^k       & 0 \\
    0      & \lambda_+^k
\end{bmatrix}
V^{-1}
\begin{bmatrix}
    0 \\
    {{f_{\mathrm{cat}}P_{1}}{e^{-\beta G n}}}
\end{bmatrix}
\end{equation}
Expanding this expression and taking the geometric sum yields $P_{n}$:
\begin{equation}
P_{n}= P_1e^{-\beta G(n-1)} {{f_{\mathrm{cat}}}\over{f_{\mathrm{cat}}-\alpha (e^{\beta G}-1)}} + A_1 \lambda_-^{n-1} + A_2 \lambda_+ ^{n-1},
\end{equation}
where the $A_1$ and $A_2$ are explicit functions of the elementary parameters. For nonzero $f_{\mathrm{cat}}$ the probability monotonically decreases for larger $n$, thus the coefficient $A_2$ must be zero. Solving this boundary condition for $I_1$ and substituting into the expression for $A_1$, we obtain the length distribution (Eq. 1 in the main text):

\begin{equation}
P_{n}= {{f_{\mathrm{cat}}}\over{f_{\mathrm{cat}}-\alpha (e^{\beta G}-1)}}P_1e^{-\beta G(n-1)}+{{\alpha(e^{\beta G}-1)}\over{\alpha (e^{\beta G}-1)-f_{\mathrm{cat}}}}P_1 e^{-D(n-1)},
\end{equation}
where $D=-\ln{\lambda_-}$.
For microtubule assembly, the physiologically relevant parameters were obtained from Ref. 21.
From the expression for $D$, we can see that the mean length and the sensitivity of the mean length to $f_{\mathrm{cat}}$ is maximal in the limit of vanishing (visualized in Fig. 1c). Expanding the expression for $D$ to first order in this limit, we obtain:
\begin{equation}
D \approx -\ln{\Bigg[1-{{f_{\mathrm{cat}}}\over{|f_{\mathrm{cat}}+\alpha+k_f-k_r|}}\Bigg]} 
\end{equation}
In this limit, the mean length retains linear sensitivity to $f_{\mathrm{cat}}$ (that is, the linear approximation to the logarithm is valid) if $\alpha > k_r-k_f + \Delta$, where the minimum buffer $\Delta$ is set by the value of $f_{\mathrm{cat}}$ because ${{f_{\mathrm{cat}}}\over{f_{\mathrm{cat}}+\Delta}}$ must be much less than 1. Therefore, as stated in the main text, the transition from weak (logarithmic) to strong (linear) catalytic regulation occurs when $\alpha > k_r - k_f$, with the sharpness being inversely proportional to $f_{\mathrm{cat}}$.

\subsection{Kinetic proofreading}

At steady state, the voltage equations taken between state 1 and the reset state in the two lower simple loops in Fig.4b are:

\begin{equation}
P_1 e^{\beta G_1} - P_{\mathrm{reset}}e^{\beta G_{\mathrm{reset}}}(1+\alpha_{\mathrm{reset}}/k_{\mathrm{reset}}) = -I_1 R_3+I_2 R_4
\end{equation}

\begin{equation}
P_1 e^{\beta G_1} - P_{\mathrm{reset,error}}e^{\beta G_{\mathrm{reset}}}(1+\alpha_{\mathrm{reset}}/k_{\mathrm{reset}}) = -I_3 R_3+I_4 R_4 e^{\beta \Delta}
\end{equation}
Setting the driven reset rate much faster than the other processes ($k_{\mathrm{reset}}/\alpha_{\mathrm{reset}} \equiv \delta << 1$), making use of the definition of the voltage drop ($\epsilon \equiv \alpha/k P e^{\beta G}$), and defining without loss of generality $G_1 \equiv 0$, these simplify to: 
\begin{equation}
P_1 - \epsilon_{\mathrm{reset}}/\delta = -I_1 R_3+I_2 R_4
\end{equation}

\begin{equation}
P_1 - \epsilon_{\mathrm{reset,error}}/\delta = -I_3 R_3+I_4 R_4 e^{\beta \Delta}
\end{equation}
The two close-loop voltage equations taken around the two upper simple loops in Fig.4b are:
\begin{equation}
(I_1+I_2)(R_1+R_2) + I_1 R_3 = P_1 (\alpha/k_1)
\end{equation}
\begin{equation}
(I_3+I_4)(R_1+R_2 e^{\beta \Delta}) + I_3 R_3 = P_1 (\alpha/k_1)
\end{equation}

Where the currents are mesh currents. These four equations can be solved, retaining the lowest order in $\delta$ to give the four steady state mesh currents:
\begin{equation}
I_1= - {{P_1(R_1+R_2-{{\alpha}\over{k_1}}R_4)}\over{R_3 R_4 + (R_1+R_2)(R_3+R_4)}}
\end{equation}
\begin{equation}
I_2= {{P_1(R_1+R_2+R_3+{{\alpha}\over{k_1}}R_3)}\over{R_3 R_4 + (R_1+R_2)(R_3+R_4)}}
\end{equation}

\begin{equation}
I_3= - {{P_1(R_1+e^{\beta \Delta}(R_2-{{\alpha}\over{k_1}}R_4))}\over{R_3 R_4 e^{\beta \Delta} + (R_1+R_2 e^{\beta \Delta})(R_3+R_4 e^{\beta \Delta})}}
\end{equation}

\begin{equation}
I_4=  {{P_1(R_1+e^{\beta \Delta}R_2+R_3(1+{{\alpha}\over{k_1}}))}\over{R_3 R_4 e^{\beta \Delta} + (R_1+R_2 e^{\beta \Delta})(R_3+R_4 e^{\beta \Delta})}}
\end{equation}
As a function of these currents, the speed, efficiency, and error are:
\begin{equation}
\mathrm{speed}=  I_2
\end{equation}
\begin{equation}
\eta =  {{I_2}\over{I_1+I_2+I_3+I_4}} 
\end{equation}

\begin{equation}
\epsilon =  I_4 /I_2 
\end{equation}

which can be combined to give:
\begin{equation}
\eta =  \Bigg(1+\epsilon + {{I_1+I_3}\over{I_2}}\Bigg)^{-1}
\end{equation}

In the biologically relevant strongly-driven regime ($\alpha/k_1 >>1$), we can further simplify to consider only the highest order in $\alpha/k_1$, giving:

\begin{equation}
\epsilon_{\mathrm{strong}} =  {{(1+{{R_2}\over{R_1}}) \eta^{-1} }\over{(e^{-\beta \Delta}+{{R_2}\over{R_1}})\eta ^{-1} -\Big( {{2e^{-\beta \Delta}}\over{1+e^{-\beta \Delta}}}+ {{R_2}\over{R_1}}\Big)(1-e^{-2\beta \Delta})}} e^{-2\beta \Delta}
\end{equation}
Note that the function ${{2x}\over{1+x}} \geq x$ for $0 \geq x \geq 1$. Since $e^{-\beta \Delta} < 1$ (the correct binding is favored over the incorrect one), this means that $\Big( {{2e^{-\beta \Delta}}\over{1+e^{-\beta \Delta}}}+ {{R_2}\over{R_1}}\Big) \geq (e^{-\beta \Delta}+{{R_2}\over{R_1}})$. Consequently,
\begin{equation}
\epsilon_{\mathrm{strong}} >  {{(1+{{R_2}\over{R_1}}) \eta^{-1} }\over{(e^{-\beta \Delta}+{{R_2}\over{R_1}})\eta^{-1} -( e^{-\beta \Delta}+ {{R_2}\over{R_1}})(1-e^{-2\beta \Delta})}} e^{-2\beta \Delta}
\end{equation}
Again making use of the fact that $e^{-\beta \Delta} < 1$, we can replace the numerator by $(e^{-\beta \Delta}+{{R_2}\over{R_1}})$ and divide out this term from the numerator and denominator to simplify to:
\begin{equation}
\epsilon_{\mathrm{strong}} >  {{\eta^{-1} }\over{\eta^{-1} -(1-e^{-2\beta \Delta})}} e^{-2\beta \Delta}.
\end{equation}
Because $e^{-2\beta \Delta}$ is typically much smaller than 1 (on the order of 0.0001), this further simplifies to Eq. 3 in the main text:
\begin{equation}
\epsilon_{\mathrm{strong}} >  {{1 }\over{1 -\eta}} e^{-2\beta \Delta}.
\end{equation}
We can see that the bound is tight, with $\epsilon$ approaching the bound for large $R_2/R_1$.
\par
Experimental values of $\epsilon$ and $\eta$ are obtained from Ref.25. To transform the catalytic efficiency to $\eta$, the measured catalytic efficiency was divided by the experimentally-inferred maximum achievable catalytic efficiency. The former is proportional to $I_2$, whereas the latter is proportional to $I_1+I_2$, which is the configuration in which all correct binding leads to successful incorporation. Because $e^{\beta \Delta} >>1$, $I_1+I_2 \approx I_1+I_2+I_3+I_4$. Therefore, $\eta$ is well approximated by the ratio of the catalytic efficiency divided by the maximum inferred catalytic efficiency.   
\par
In the weakly-driven regime ($\alpha/k_1 <<1$), the expression simplifies to:
\begin{equation}
\begin{split}
\epsilon_{\mathrm{weak}} =  {{(R_1+e^{\beta \Delta}R_2+R_3)(R_3R_4+(R_1+R_2)(R_3+R_4)) }\over{(R_1+R_2+R_3) \Big( R_1(R_3+e^{\beta \Delta}R_4)+e^{\beta \Delta}(R_3R_4+R_2(R_3+e^{\beta \Delta}R_4))\Big)}} \\ + {{(1-e^{\beta \Delta})R_2(R_3R_4+(R_1+R_2)(R_3+R_4)) \eta^{-1}\ln{[1+\alpha/k_1]} }\over{(R_1+R_2+R_3) \Big( R_3 R_4(1+e^{\beta \Delta})+R_1(2R_3+R_4+e^{\beta \Delta}R_4)+R_2(R_3(1+e^{\beta \Delta})+R_4(1+e^{2\beta\Delta}))\Big)}}
\end{split}
\end{equation}
Because there is an extra factor of $e^{\beta \Delta}$ in the denominators of both terms, the minimum value of $\epsilon_{\mathrm{weak}}$ is obtained in the limit that $e^{\beta \Delta}$ becomes much larger than any ratio of resistors. Therefore, keeping only the highest order in $e^{\beta \Delta}$ yields a lower bound on the efficiency which is tight in the limit of large $\Delta$:
\begin{equation}
\epsilon_{\mathrm{weak}} >  {{(1- \eta^{-1}\ln{[1+\alpha/k_1]} )\Big(R_3(R_1+R_2)+R_4(R_1+R_2+R_3) \Big) }\over{R_4(R_1+R_2+R_3)}} e^{-\beta \Delta},
\end{equation}
which further simplifies to:
\begin{equation}
\epsilon_{\mathrm{weak}} >  (1-\eta^{-1}\ln{[1+\alpha/k_1]} ) e^{-\beta \Delta},
\end{equation}
which is tight if $R_4 >> R_3$. In the weakly-driven limit, $\ln{[1+\alpha/k_1]}  \approx \alpha/k_1 << 1$. Consequently, $1-\eta^{-1}\ln{[1+\alpha/k_1]} \approx e^{-\eta^{-1}\ln{[1+\alpha/k_1]} }$, and we obtain the tight error bound in the weakly-driven limit:
\begin{equation}
\epsilon_{\mathrm{weak}} >  e^{-{{\alpha}\over{k_1\eta}}} e^{-\beta \Delta}.
\end{equation}

\end{widetext}

\end{document}